\documentclass[prd,showpacs,preprintnumbers,amsmath,amssymb,floatfix]{revtex4}

\usepackage{graphicx}
\usepackage{dcolumn}
\usepackage{bm}
\usepackage{amssymb}
\usepackage{epsfig}
\usepackage{color}

\newcommand{\ba}{\begin{eqnarray}}
\newcommand{\ea}{\end{eqnarray}}
\newcommand{\be}{\begin{equation}}
\newcommand{\ee}{\end{equation}}
\newcommand{\bdisplay}{\begin{displaymath}}
\newcommand{\edisplay}{\end{displaymath}}

\newcommand{\eq}[1]{Eq.\,(\ref{#1})}
\newcommand{\fig}[1]{Fig.\,\ref{#1}}

\begin{document}

\title{The slope, curvature, and higher parameters in $pp$ and $\bar{p}p$ scattering, and the extrapolation of measurements of $d\sigma(s,t)/dt$ to $t=0$}

\author{Martin~M.~Block}
\email{mblock@northwestern.edu}
\affiliation{Department of Physics and Astronomy, Northwestern University,
Evanston, IL 60208}
\author{Loyal Durand}
\email{ldurand@hep.wisc.edu}
\altaffiliation{Mailing address: 415 Pearl Ct., Aspen, CO 81611}
\affiliation{Department of Physics, University of Wisconsin, Madison, WI 53706}
\author{Phuoc Ha}
\email{pdha@towson.edu}
\affiliation{Department of Physics, Astronomy and Geosciences, Towson University, Towson, MD 21252}
\author{Francis Halzen}
\email{francis.halzen@icecube.wisc.edu}
\affiliation{Wisconsin IceCube Particle Astrophysics Center and Department
of Physics, University of Wisconsin-Madison,  Madison,  Wisconsin 53706}

\begin{abstract}
We study the effects of curvature in the expansion of the logarithm of the differential elastic scattering cross section  near $t=0$ as $d\sigma(s,t)/dt=d\sigma(s,0)/dt\,\times\exp(Bt+Ct^2+Dt^3\cdots)$ in an eikonal model for $pp$ and $\bar{p}p$ scattering, and use the results to discuss the extrapolation of measured differential cross sections and the slope parameters $B$ to $t=-q^2=0$. We find that the curvature effects represented by the parameters $C$ and $D$, while small, lead to significant changes in the forward slope parameter relative to that determined in a purely exponential fit, and to smaller but still significant changes in the forward elastic scattering and total cross sections. Curvature effects should therefore be considered in future analyses or reanalyses of the elastic scattering data.
\end{abstract}

\pacs{13.85.Dz, 13.85.Lg}

\maketitle

%%%%%%%%%%%%%%%%%%%%%%%%%
%%%%%%%%%%%%%%%%%%%%%%%%%

\section{ Background \label{sec:background}}

 The differential elastic scattering cross sections $d\sigma/dt$ in  proton-proton $(pp)$  or antiproton-proton $(\bar{p}p)$ elastic scattering  are generally described as decreasing purely exponentially at  small values of momentum transfer variable $t$ in the scattering, with $d\sigma/dt\approx A\exp(Bt)=A\exp(-B|t|)$. This general behavior is evident in the semi-logarithmic plots usually used to display the data at small angles or low $|t|$, with $d\sigma/dt$ decreasing nearly linearly in those plots until $|t|$ approaches  the first diffraction minimum seen in the high energy cross sections.

 This behavior is used to extrapolate the differential cross section to $t=0$ to determine the forward slope parameter $B$, the forward elastic scattering cross section, and indirectly through that, the total cross section $\sigma_{\rm tot}$. Any curvature in $\ln(d\sigma/dt)$ at small $t$ is clearly small at high energies, but if present could affect the extrapolation and the determination of those quantities.

In a recent analysis of its results on $pp$ scattering at 8 TeV at the Large Hadron Collider (LHC) \cite{totem2015}, the TOTEM Collaboration established that the differential elastic scattering cross section  does not, in fact, decrease purely exponentially as $\exp(-B|t|)$ at small values of $t$, but rather shows a positive upward curvature relative to the expected exponential decrease as $|t|$ increases from zero. Curvature has been studied previously at much lower energies \cite{blockcahn}, but has generally not been considered in most analyses of high energy scattering at small momentum transfers.

Our objective in this paper is to present a careful study of curvature effects in the differential cross sections using the comprehensive eikonal fit  in \cite{bdhh-eikonal} to the total, elastic, and inelastic $pp$ and $\bar{p}p$ cross sections from 10 GeV to 57 TeV, the measured slope parameters $B$, and the ratios $\rho$ of the real to the imaginary parts of the forward scattering amplitudes in that energy range, and to apply the results to the analysis of a representative set of high energy experiments. We find that the effects of curvature are generally small, as expected, but change the values of the forward differential cross sections and the forward slope parameters by amounts that are significant on the scale of the quoted experimental uncertainties.  Curvature effects should therefore be taken into account in future analyses or reanalyses of the elastic scattering data.

We note in this context that the detailed study of hadronic scattering amplitudes at small momentum transfers has a long history, and provides a potential window on new physics. Thresholds associated with the appearance of new phenomena, for example the opening of new spatial dimensions or the onset of new types of interactions, leave an imprint on the energy dependence of the forward scattering amplitudes that determine the total, elastic, and inelastic hadronic cross sections. While recent measurements at the LHC have convincingly reinforced earlier evidence that the $pp$ and $\bar{p}p$ scattering amplitudes asymptotically approach those for black-disk scattering \cite{bdhhfit} and provide no evidence so far of new thresholds, detailed analysis of higher energy and higher precision data, with reliable extrapolation of the scattering amplitudes to $t=0$, should be of great interest.

According to the treatment of the effects of Coulomb scattering at small momentum transfers by West and Yennie \cite{west-yennie} and Cahn \cite{cahn,blockcahn}, we can write the  differential elastic scattering amplitude for $pp$ (or $\bar{p}p$) scattering  amplitude as the sum of Coulomb and hadronic parts, with the Coulomb part known. We are primarily interested in the hadronic or strong-interaction part of the scattering, but will include the Coulomb effects in our later analyses of high energy experiments. For simplicity, we concentrate now on the purely hadronic effects and take $f(s,t)$ as the hadronic part of the scattering amplitude.

We begin with the general expression for this part of the scattering amplitude written in an impact parameter representation,
\be
\label{f}
f(s,t) = i\int_0^\infty db\,b\left(1-e^{i\chi(s,b)}\right) J_0(b\sqrt{-t}).
\ee
The hadronic part of the differential elastic scattering cross section is then
\be
\label{dsigma/dt}
\frac{d\sigma}{dt}(s,t)= \pi\left|f(s,t)\right|^2.
\ee
Here $s=W^2=4(p^2+m^2)$ is the square of the total energy  in the center of mass (c.m.) system, $p$ is the c.m. momentum of either incident particle, $b=\sqrt{j(j+1)}\big/p$ where $j$ is the partial-wave angular momentum, and $t=-2p^2(1-\cos\theta)$ is the invariant 4-momentum transfer for elastic scattering at the angle $\theta$.

We will write the complex eikonal function $\chi(s,b)$  as $\chi=\chi_R+i\chi_I$; note that some other papers use different conventions, {\em e.g.,} \cite{edge,blockrev}. With this convention, the total hadronic scattering cross section is
\be
\label{sigma_tot}
\sigma_{\rm tot}(s)= 4\pi {\rm Im} f(s,0) = 4\pi \int_0^\infty db\, b \left (1-\cos{\chi_R}\,e^{-\chi_I}\right).
\ee

We will assume that the momentum-transfer dependence of the differential cross section $d\sigma(s,t)/dt\equiv d\sigma(W,q^2)/dq^2$ at the center-of-mass energy $W=\sqrt{s}$ can be parametrized for $q^2=-t$ near zero as an exponential,
\ba
\label{tseries}
\frac{d\sigma}{dt}(W,t) &\approx& \frac{d\sigma}{dt}(W,0)e^{B t+C t^2 + D t^3+\cdots} \\
\label{CDdefined}
&=& \frac{d\sigma}{dq^2}(W,0)e^{-B q^2+C q^4-D q^6+\cdots}
\ea
where, in \eq{CDdefined}, we have switched from $t$ to $q^2=|t|=2p^2(1-\cos{\theta})$ as the variable to eliminate the overall sign of $t$ in the physical region. $B$ is the forward slope parameter. We will call $C$, $D$, and higher order terms collectively `curvature parameters,' and will also parametrize the forward cross section $d\sigma/dq^2|_{q^2=0}$ (in mb/GeV$^2$) as $\exp{(A)}$. Then $\ln(d\sigma/dq^2)=A-Bq^2+Cq^4+\cdots$.

Unfortunately, the values of $d\sigma(W,0)/dq^2$, $B$, $C$, and the higher derivatives of $\ln(d\sigma/dq^2)$ at $q^2=0$ are not directly accessible in experiment: the values quoted by experimenters are obtained by fitting data on $d\sigma/dq^2$ over a range of $q^2$ near zero using an expression of the form in \eq{CDdefined}.  It is not immediately clear how many terms should be included in the exponent, or how the range of $t$ or $q^2$  should be restricted to get reliable results. It is common, in fact, to assume that $d\sigma/dq^2$ is purely exponential at small $q^2$ and use the fitted result to determine $B$ and $d\sigma(W,0)/dq^2$. We therefore turn here to the calculation of the slope, curvature, and other parameters at non-zero values of $q^2$ where they can be determined directly, and then use these theoretical results to determine how fits to experiment of the type above can be extrapolated $q^2=0$.

The presence of curvature in $d\sigma/dq^2$ at small $q^2$ has been established at 8 TeV by the TOTEM group \cite{totem2015}. It had already been seen at lower energies---see \cite{blockcahn} for a discussion.

%%%%%%%%%%%%%%%%%%%%%%
%%%%%%%%%%%%%%%%%%%%%%

\section{Derivation of the differential parameters \label{sec:derivation}}

To obtain series expansions for the logarithm of $d\sigma/dq^2$ about a point $q_0^2\geq 0$, we start with the general expression
for the complex scattering amplitude $f(W,q^2) = f_R(w,q^2)+if_I(W,q^2)$,
\ba
\label{fR}
f_R(W,q^2) &=& -\int_0^\infty db\,\sin{\chi_R} e^{-\chi_I} J_0(qb), \\
\label{fI}
f_I(W,q^2) &=& \int_0^\infty db\,(1-\cos{\chi_R} e^{-\chi_I}) J_0(qb),
\ea
and expand the Bessel functions in Taylor series about the point $q=q_0$ using the expression
\be
\label{Bessel_series}
J_0(qb) = \sum_{k=0}^\infty \frac{(-1)^k}{2^k k!} (q^2-q_0^2)^k J_k(q_0b)/(q_0b)^k.
\ee

We next define the amplitudes
\ba
\label{mkR}
m_{k,R}(W,q_0) &=& -2^k k! \int_0^\infty db\,b^{2k+1}\sin{\chi_R} e^{-\chi_I} J_k(q_0b)/(q_0b)^k, \\
\label{mkI}
m_{k,I}(W,q_0) &=& 2^k k! \int_0^\infty db\,b^{2k+1}(1-\cos{\chi_R} e^{-\chi_I})  J_k(q_0b)/(q_0b)^k,
\ea
With these definitions,
\be
\label{dsig}
\frac{d\sigma}{dq^2}(W,q^2) = \left(\sum_{k=0}^\infty \frac{(-1)^k}{(2^kk!)^2}  (q^2-q_0^2)^k m_{k,I}\right)^2 +  \left(\sum_{k=0}^\infty \frac{(-1)^k}{(2^kk!)^2}  (q^2-q_0^2)^k m_{k,R}\right)^2.
\ee

We note that $d\sigma(W,q_0^2)/dq^2=m_{0,I}^2+m_{0,R}^2$, extract this factor from the expression in \eq{dsig}, and write
\ba
\label{dsig_exp_form}
\frac{d\sigma}{dq^2}(W,q^2) &=& \frac{d\sigma}{dq^2}(W,q_0^2)\exp\left\{\log\left[\frac{d\sigma}{dq^2}(W,q^2) \bigg/\frac{d\sigma}{dq^2}(W,q_0^2)\right]\right\} \\
\label{dsig_exp_form2}
&=& \frac{d\sigma}{dq^2}(W,q_0^2)\exp\left\{\log\left[\left(\sum_{k=0}^\infty \frac{(-1)^k}{(2^kk!)^2} (q^2-q_0^2)^k M_k^I\right)^2 +  \left(\sum_{k=0}^\infty \frac{(-1)^k}{(2^kk!)^2} (q^2-q_0^2)^k M_k^R\right)^2\right]\right\},
\ea
where $M_k^I$ and $M_k^R$ are the normalized integrals
 \ba
 \label{M_kI}
M_k^I &=& m_{k,I}\big/(m_{0,I}^2+m_{0,R}^2)^{1/2}, \\
\label{M_kR}
M_k^R &=& m_{k,R}\big/(m_{0,I}^2+m_{0,R}^2)^{1/2}.
\ea
The exponential factor in \eq{dsig_exp_form2} describes the behavior of the differential cross section near $q^2=q_0^2$ and gives the parameters $B,\ C, \cdots$ for $q_0^2=0$.

The leading term in the expression in square brackets in \eq{dsig_exp_form2} arises from $k=0$ in the sums; this is just $(M_0^I)^2+(M_0^R)^2=1$. The remaining terms are small for $q^2\approx q_0^2$. Expanding the logarithm for $|q^2-q_0^2|<<1$, we find that
\ba
\frac{d\sigma}{dt}(W,q^2) &=& \frac{d\sigma}{dq^2}(W,q_0^2)\exp\left\{-\frac{1}{2}\left(M_0^I M_1^I+M_0^R M_1^R\right)\left(q^2-q_0^2\right)\right. \nonumber \\
&&  +\left[-\frac{1}{8}\big(M_0^I M_1^I+M_0^R M_1^R\big)^2+\frac{1}{32}\big(2( M_1^I)^2+2( M_1^R)^2+M_0^I M_2^I+M_0^R M_2^R\big)\right] (q^2-q_0^2)^2 \nonumber \\
\label{dsig_expanded}
&& +\frac{1}{1152}\left[ -\frac{1}{24}\big(M_0^I M_1^I+M_0^R M_1^R\big)^3 +\frac{1}{64}\left(M_0^I M_1^I+M_0^R M_1^R\right) \big(2( M_1^I)^2+2( M_1^R)^2+M_0^I M_2^I+M_0^R M_2^R\big)  \right. \nonumber \\
&& -\big(9 M_1^I M_2^I+9M_1^R M_2^R+M_0^I M_3^I+M_0^R M_3^R\big)\bigg](q^2-q_0^2)^3+\cdots\bigg\}.
\ea
The argument of the exponential in \eq{dsig_expanded} gives the expansion of the logarithm of the ratio of $d\sigma/dq^2$ at $q^2$ to its value at $q_0^2$, hence, the expressions we want for the logarithmic slope, curvature, and higher coefficients at $q_0^2$. In particular,
\ba
\label{B(q0^2)}
B(q_0^2) &=& \frac{1}{2}\left(M_0^I M_1^I+M_0^R M_1^R\right), \\
\label{C(q0^2)}
C(q_0^2) &=& -\frac{1}{8}\big(M_0^I M_1^I+M_0^R M_1^R\big)^2+\frac{1}{32}\big(2( M_1^I)^2+2( M_1^R)^2+M_0^I M_2^I+M_0^R M_2^R\big), \\
\label{D(q0^2)}
D(q_0^2) &=& \frac{1}{1152}\left[ \frac{1}{24}\big(M_0^I M_1^I+M_0^R M_1^R\big)^3 -\frac{1}{64}\left(M_0^I M_1^I+M_0^R M_1^R\right) \big(2( M_1^I)^2+2( M_1^R)^2+M_0^I M_2^I+M_0^R M_2^R\big)  \right. \nonumber \\
&& +\big(9 M_1^I M_2^I+9M_1^R M_2^R+M_0^I M_3^I+M_0^R M_3^R\big)\bigg].
\ea

These expression simplify considerably if the real part of the scattering amplitude is small enough to neglect. Then $M_0^I=1$ and all the real part terms $M_k^R$ vanish.

%%%%%%%%%%%%%%%%%%%%%%
%%%%%%%%%%%%%%%%%%%%%%

\section{Behavior of $B$, $C$, and $D$ for $pp$ and $\bar{p}p$ scattering and their effect on the cross sections \label{sec:BCD}}

%%%%%%%%%%%%%%%%%%%%%%

We have used the detailed eikonal model of high-energy $pp$ and $\bar{p}p$ scattering we discussed in \cite{bdhh-eikonal} to study the behavior and importance of the differential parameters $B$, $C$, and $D$ in those processes. The model provides a comprehensive description of the $pp$ and $\bar{p}p$ scattering cross sections $\sigma_{\rm tot}$, $\sigma_{\rm elas}$, $\sigma_{\rm inel}$, and $d\sigma/dq^2$, the ratios $\rho={\rm Re}\,f(s,0)/{\rm Im}\,f(s,0)$ of the real to the imaginary parts of the forward scattering amplitudes, and the slope parameters $B$ as determined in experiments over the energy range 10 GeV to 57 TeV.  Our fit to the extensive high-energy data is very good, with a $\chi^2$ of 173 for 157 degrees of freedom; we believe it is sufficiently accurate for present purposes.

\begin{figure}[htbp]
\includegraphics{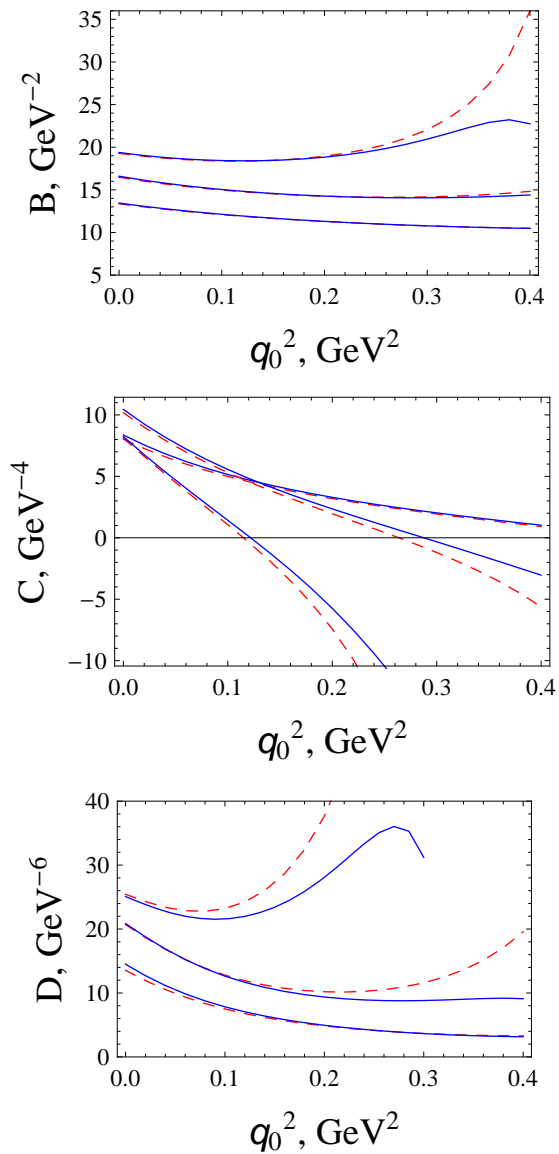}
\caption{Plots of the dependence of the slope and curvature parameters $B(q_0^2)$, $C(q_0^2)$ and $D(q_0^2)$ on the  local momentum transfer $q_0^2$ for center-of-mass energies $W=7000,\, 1000$ and 100 GeV (curves top to bottom for $B$ and $D$, and bottom to top for $C$ with $q_0^2>0.2$). The full results are given by the solid (blue) curves. The results obtained neglecting the contributions of the real part of the scattering amplitude are given by the dashed (red) curves.  }
\label{fig:BCDcomp}
\end{figure}

We show our calculated values of $B$, $C$, and $D$ in \fig{fig:BCDcomp}  as functions of the local momentum transfer $q_0^2$ for center-of-mass energies of 100, 1000, and 7000 GeV. We also show the results we obtain neglecting the contributions of the real part of the scattering amplitude in Eqs.\ (\ref{B(q0^2)})-(\ref{D(q0^2)}) which leads to considerable simplifications. While  this  is a reasonable approximation for determining the overall behavior of $B$, $C$, and $D$ at small values of  $q_0^2$,  we find that even  the small errors in $B$   are on the order of the experimental uncertainties in that quantity. We will therefore use the full expressions in what follows.

The effects of the real part on all the parameters grow at larger values of $q_0^2$ close to the first diffraction zero in the imaginary part of the scattering amplitude. The approach of the diffraction zero is especially evident in the local curvature $C$, which changes sign from positive to negative at decreasing values of $q_0^2$ as the energy increases. This effect was noted in \cite{blockcahn}, where the change in sign of $C$ was taken as a sign of the approach to the black-disk limit of the scattering.

%%%%%%%%%%%%%%%%%%%%%
%%%%%% FIG 1 BCD  %%%%%

%%%%%%%%%%%%%%%%%%%%%
%

In order to extract the forward slope parameter $B\equiv B(0)$ and the forward differential cross section $d\sigma/dq^2\big|_{q^2=0}$ from measured differential cross sections, experimenters  typically analyze  their data using the simple exponential form
\be
\label{exp_approx}
\frac{d\sigma}{dq^2}(W,q^2)\approx \frac{d\sigma}{dq^2}(W,0)e^{-Bq^2} \equiv e^{A-Bq^2}
\ee
to describe the purely hadronic part of the scattering, plus additional terms which describe the effects of Coulomb scattering and Coulomb-hadronic interference \cite{west-yennie,cahn,blockcahn}. The forward slope parameter $B$ and cross section $d\sigma/dq^2\big|_{q^2=0}= e^A$ are then determined by fitting the data on $d\sigma/dq^2$ over ranges of $q^2$ as close as possible to the  forward direction $q^2=0$.

\begin{figure}[htbp]
\includegraphics{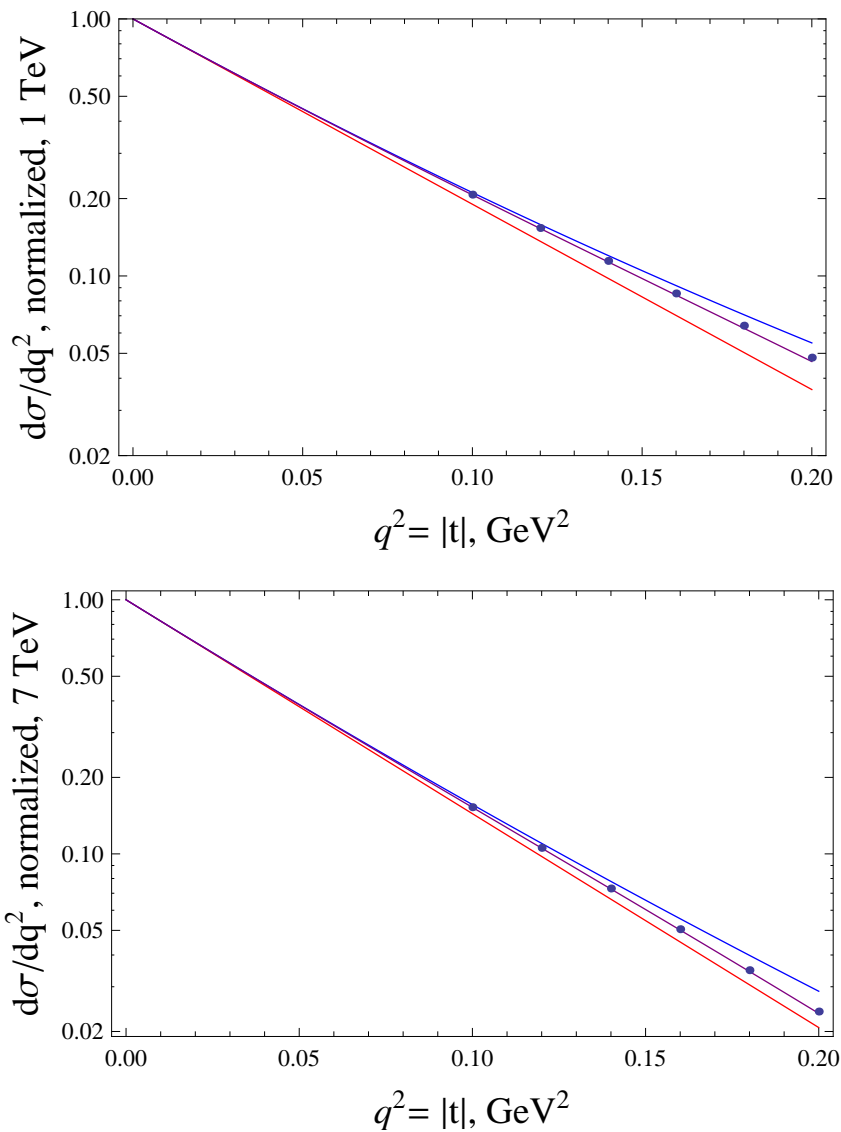}
\caption{ Plots of the ratios of the cross sections $d\sigma/dq^2$ to the forward cross sections at 1000 GeV and 7000 GeV obtained in the  approximations of purely exponential behavior $\exp(-B q^2)$ near $q^2=0$ (bottom red curves), and including the additional terms $+Cq^4$ (top blue curves) and $-Dq^6$ (central purple curves) in the expansion of the exponent using the values of $B$, $C$, and $D$ at $q^2=0$ obtained in the eikonal fit to the high-energy $pp$ and $\bar{p}p$ data in \cite{bdhh-eikonal}. The exact results for the ratios are shown as black dots. }
\label{fig:SeriesExactComp}
\end{figure}

A question immediately arises as to the accuracy of this procedure: the effective values of $A$ and $B$ presumably correspond to  local values at a $q_0^2$ somewhere near the middle of the experimental interval, not $q^2=0$, and the possible effects of the curvature terms are ignored. In \fig{fig:SeriesExactComp} we show the effects of the latter as obtained in the eikonal model of \cite{bdhh-eikonal} at $W=1000$ and 7000 GeV, regions of considerable experimental interest. The results at lower energies are similar.

The curves in \fig{fig:SeriesExactComp} show the ratios
\be
\label{SeriesExactRatios}
\frac{d\sigma}{dq^2}(W,q^2)\Big/\frac{d\sigma}{dq^2}(W,0)=e^{-Bq^2+Cq^4-Dq^6+\cdots}
\ee
in the successive approximations of including only the $B$ term, the $B$ and $C$ terms, and the $B$, $C$, and $D$ terms, compared to the exact results of the model. The individual curves seem, over limited ranges of $q^2$ in a semi-logarithmic plot, to be nearly exponential, but the effects of the curvature terms are clearly important since the local slopes differ noticeably from the constant forward slope $B$. For reference,  the ranges of $q^2$ used in the analyses of the TOTEM data at 8000 GeV \cite{totem2015}, the ATLAS data at 7000 GeV \cite{atlas2014}, and the E710 data at 1800 GeV \cite{E710_B2,E710lumin-ind,E710pbar-p-final} are $q^2=0.029$--0.19 GeV$^2$, 0.01--0.1 GeV$^2$, and 0.02--0.08 GeV$^2$, ranges for which the deviations of the apparent slopes from $B(0)$ are noticeable.

%%%%%%%%%%%%%%%%%%%%%
%%%%%% FIG 2  %%%%%

%%%%%%%%%%%%%%%%%%%%%
%

\begin{figure}[htbp]
\includegraphics{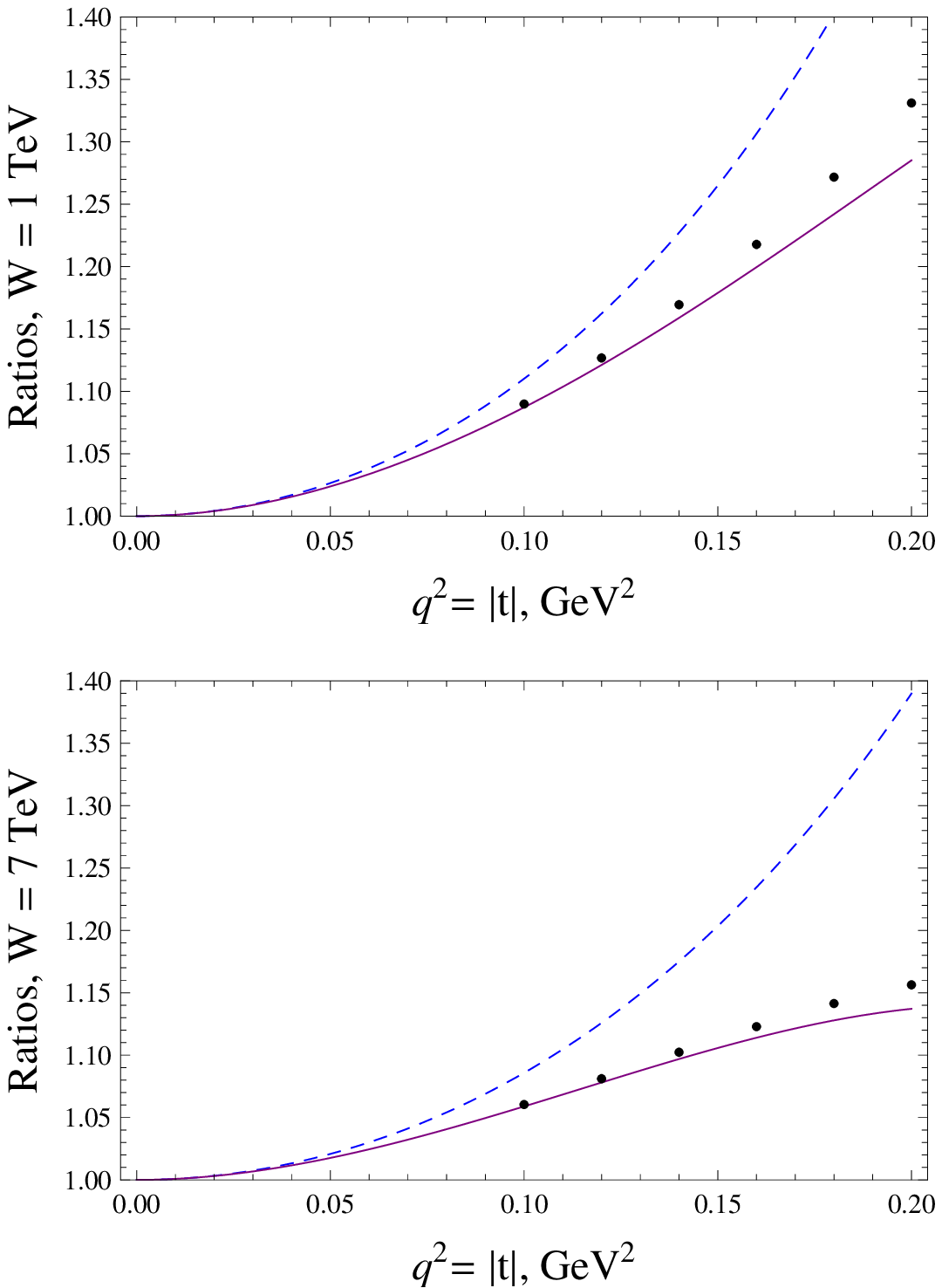}
\caption{ Plots of the calculated ratios of the cross sections $d\sigma/dq^2$  near $q^2=|t|=0$  to their exponential approximations $d\sigma/dq^2\big|_{q^2=0}\times\, \exp(-Bq^2)$ (black dots) at 1000 and 7000 GeV compared to the leading approximations for those ratios, $\exp(+Cq^4)$ (dashed blue curves) and $\exp(+Cq^4-Dq^6)$ (solid purple curves), in an expansion of $\ln(d\sigma/dq^2)$ about $q^2=0$. The cross sections and the expansion coefficients $B$, $C$ and $D$ were  obtained using the eikonal fit to the high-energy $pp$ and $\bar{p}p$ data in \cite{bdhh-eikonal}.  }
\label{fig:SeriesRatioComp}
\end{figure}

We show these effects in a different way in \fig{fig:SeriesRatioComp} where we plot the same ratios of cross sections, but with the dominant, exponentially decreasing factor $e^{-Bq^2}$ divided out. The figure shows clearly the relatively large corrections to the simple exponential form of the cross section associated with the curvature term $C$, and with $C$ plus $D$. The size of these corrections is suppressed in a standard semi-logarithmic plot of the differential cross section such as \fig{fig:SeriesExactComp}.

%%%%%%%%%%%%%%%%%%%%%
%%%%%% FIG 3  %%%%%

%%%%%%%%%%%%%%%%%%%%%
%

The $C$ and $D$ (and higher) terms affect the local slope of $\ln(d\sigma/dq^2)$. In terms of the series expansion of $\ln(d\sigma/dq^2)$ around $q^2=0$,
\be
\label{Blocal}
B(q_0^2) = -\frac{d}{dq^2}\ln(d\sigma/dq^2)\big|_{q^2=q_0^2} = B-2Cq_0^2+3Dq_0^4+\cdots.
\ee
The local slope is just the tangent to the cross section curve in \fig{fig:SeriesExactComp} at $q_0^2$, and as such, is approximately the slope we would find in a fit to that curve over an interval  around $q_0^2$. We would then  identify $B(q_0^2)$ as $B=B(0)$ in an exponential  model for $d\sigma/dq^2$; this is the common procedure in fitting data.

The $C$ and $D$ terms in \eq{Blocal}  give the approximate amount by which we have to change the local slope  to find the forward slope $B$, $B\approx B(q_0^2)+2Cq_0^2-Dq_0^4+\cdots$. As we will see in the next section, the corrections to $B$ are small, but still significant for the ranges of $q^2$ typical in experiments at high energies.

We also see from the deviation of the exact results  in \fig{fig:SeriesRatioComp} from the curves in that figure that higher-order terms need to be included in the series expansion of  $\ln(d\sigma/dq^2)$ at the higher values of $q^2$ shown, with $\ln(d\sigma/dq^2)=A-Bq^2+Cq^4-Dq^6+Eq^8-\cdots$. However, it is not clear that this would be useful since the series apparently converges slowly. We will instead restrict the range of $q^2$ used in our analysis of experimental data in the next section to that where the deviations are small enough to ignore.

%%%%%%%%%%%%%%%%%%%%%%
%%%%%%%%%%%%%%%%%%%%%%

\section{Applications to experiment \label{sec:applications}}

%%%%%%%%%%%%%%%%%%%%%%

\subsection{Fits to the differential cross sections \label{subsec:fits-to-xsecs}}

In this section, we will apply the results above to the analysis of $pp$ scattering at the CERN Intersecting Storage Rings (ISR) at 52.8 GeV,   $\bar{p}p$ scattering in experiment E710 at the Fermilab Tevatron at 1800 GeV,  and $pp$ scattering at the Large Hadron Collider at 7000 GeV (ATLAS-ALFA) and 8000 GeV (TOTEM). We find that the effects of the curvature-type terms $C$ and $D$ on simple exponential fits change the fitted values of $B$ and the forward differential cross section by small but significant amounts, and discuss the implications for derived values of the total cross sections.  We emphasize that our analyses are based on straightforward least squares fits to the rather precise data at those energies using only the quoted statistical errors for the different experiments; definitive (re)analyses of these and other experiments will be left to the respective experimental groups.

The results in the previous section show that the effects of the $C$ and $D$ terms on the differential $pp$ and $\bar{p}p$ cross sections are significant in the range of $q^2$ typically used in analyses of high energy data, changing the calculated cross sections by up to $\approx 10\% \ (30\%)$  for $q^2=0.1\ ( 0.2)$ GeV$^2$ relative to the simple exponential form $d\sigma/dq^2=\exp{(A-B q^2)}$ with the correct values of $A$ and $B$; this is shown in \fig{fig:SeriesRatioComp}.  The problem, given an exponential fit to experimental data over some range of non-zero $q^2$, is one of extrapolation to $q^2=0$ to obtain the correct values of the forward parameters $A$ and $B$.

%%%%%%%%%%%%%%%%%%%%%
%%%%%% FIG 4  %%%%%

\begin{figure}[htbp]
\includegraphics{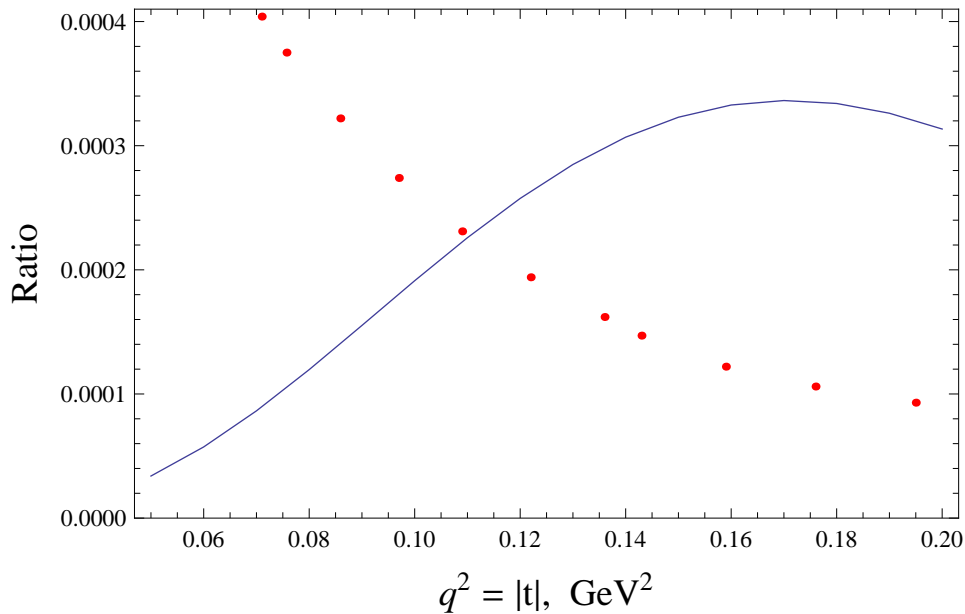}
\caption{The difference between the exact ratio of the differential cross section $d\sigma/dq^2$ at $q^2$ to its value at $q^2 = 0$,  and the approximate expression $\exp(-B q^2+Cq^4-D q^6)$ for that ratio at 8000 GeV (solid blue curve),
compared to the ratios of the experimental statistical errors in the cross section to the forward cross section for representative TOTEM points \cite{totem2015}. The cross section and the coefficients $B$, $C$, and $D$ were calculated using the eikonal fit to the high-energy $pp$ and $\bar{p}p$ data in \cite{bdhh-eikonal}.   }
\label{fig:RatioErrorPlot}
\end{figure}
%%%%%%%%%%%%%%%%%%%%%
%

  We cannot use the eikonal model of \cite{bdhh-eikonal} directly in the extrapolation of individual results. It is the result of a very good comprehensive fit to the experimental cross section data from 10 GeV to 57 TeV, but not all experimental results agree. Further, the fitted values of $B$, $\rho$, and the cross sections have themselves been determined over various, often differing, ranges of $q^2$ with the curvature terms  generally neglected, so do not correspond to their properly-extrapolated values, and we can anticipate small future changes in the eikonal.

 As shown above, the extended exponential form which includes the $C$ and $D$ terms in the expansion of $\ln(d\sigma/dq^2)$ gives an excellent fit to the exact eikonal results for $q^2$ sufficiently small, so should work quite generally;  we will use this form in the following analysis.  We find that is generally not possible to determine all four parameters $A,\,\ldots,D$ in fits to individual data sets because of the statistical limitations of the data and the smallness of the curvature corrections.  We are primarily interested in $A$ and $B$ in any case. We will therefore adopt a hybrid approach, keep $A$ and $B$ as free parameters, and take $C$ and $D$ from the eikonal fit in \cite{bdhh-eikonal}. We believe this should give reliable results given the overall success of the eikonal model and the smallness of the curvature corrections; errors in the latter should be considerably suppressed in the final results.

The results in \fig{fig:SeriesRatioComp} suggest that we should restrict our analysis to values of $q^2\lesssim 0.1$ GeV$^2$ where the deviations of our extended exponential model from the exact results for the cross sections are very small, $\lesssim 0.5\%$. We note that those deviations grow rapidly at larger $q^2$. We should also require that the deviations be small relative to the statistical errors in the cross sections we are attempting to fit so that the deviations do not bias the fit. As seen in \fig{fig:RatioErrorPlot}, this leads to essentially the same restriction, $q^2\lesssim 0.1$ GeV$^2$, in the case of the TOTEM data.\footnote{The TOTEM analysis in \cite{totem2015} used data out to $q^2\approx 0.2$ GeV$^2$.} There was no further restriction for the ISR \cite{ISR1}, E710 \cite{E710_B2,E710lumin-ind,E710pbar-p-final}, or ATLAS \cite{atlas2014} data used in our analysis. We have adopted the restriction $q^2\leq 0.1$ GeV$^2$ in our fitting procedure.

 We used the form of the hadronic cross section in \eq{CDdefined}, with $C$ and $D$ taken from our eikonal results,  to reanalyze the accurate ISR data at 52.8 GeV over the range $0.001\leq q^2\leq 0.055$ GeV$^2$ \cite{ISR1}, the E710 data at 1800 GeV over the range $0.0339\leq q^2\leq 0.103$ GeV$^2$ \cite{E710_B2,E710lumin-ind,E710pbar-p-final}, the ATLAS data 7000 GeV over the range $0.011\leq q^2\leq 0.0959$ GeV$^2$ \cite{atlas2014}, and the TOTEM data at 8000 GeV over the range $0.027 \leq q^2\leq 0.103$ GeV$^2$ \cite{totem2015}, using the statistical uncertainties quoted in those references in a least squares fit. We included the Coulomb scattering corrections and the Coulomb-hadronic interference terms \cite{cahn,blockcahn} in all cases even though they are small at the highest energies where the data do not extend to the very small values of $q^2$ necessary to see the Coulomb peak directly.

\begin{figure}[htbp]
\includegraphics{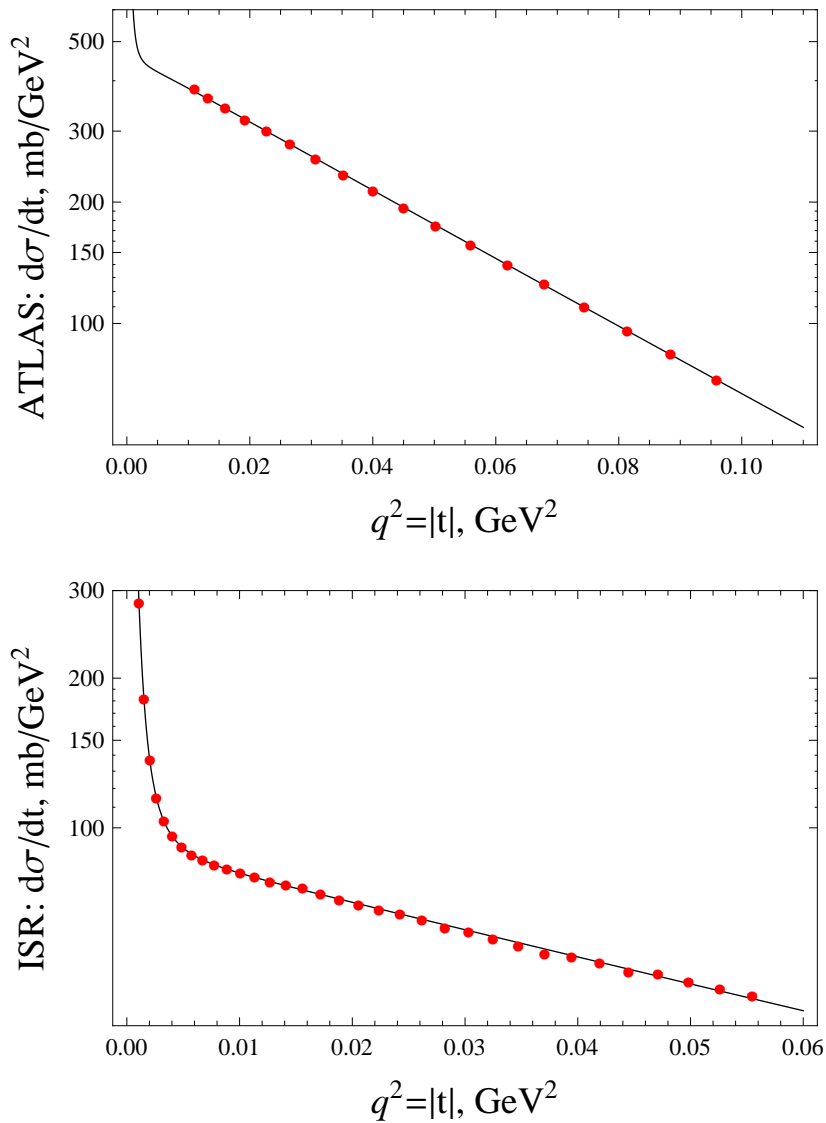}
\caption{Fits to the differential $pp$ elastic scattering cross sections $d\sigma/dq^2=d\sigma/d|t|$ for the ATLAS data at 7000 GeV \cite{atlas2014} and the ISR data at 52.8 GeV \cite{ISR1}. The values of the curvature terms $C$ and $D$ in the series expansion of the hadronic contribution to $\ln(d\sigma/dq^2$ were taken from the overall eikonal fit to the high-energy $pp$ and $\bar{p}p$ data in \cite{bdhh-eikonal}. The effects of the Coulomb interactions between the protons at small momentum transfers were included as described in the ATLAS and ISR papers. }
\label{fig:ATLAS-ISR_xsecs}
\end{figure}

 We summarize the results of our fits  in  Table I, and show the fits to $d\sigma/dq^2$ for the ATLAS data at 7000 GeV and the ISR data at 52.8 GeV in \fig{fig:ATLAS-ISR_xsecs} in conventional semi-logarithmic plots.

 %%%%%%%%%%%%%%%%%%%%%
%%%%%% FIG 5  %%%%%

%%%%%%%%%%%%%%%%%%%%%
%

 %%%%%%%%%%%%% TABLE I %%%%%%%%%%%
 %%%%%%%%%%%%%%%%%%%%%%%%%%%%%

 \begin{table}[ht]                   % Use "table" environment, but also
 \caption{\protect  The results of our fits the data of the TOTEM \cite{totem2015}, ATLAS \cite{atlas2014}, E710 \cite{E710_B2,E710lumin-ind,E710pbar-p-final}, and IRS  \cite{ISR1} experiments at 8000, 7000, 1800, and 52.8 GeV, respectively. $A_{\rm exp}$ and $B_{\rm exp}$ are the results of purely exponential fits to the hadronic part of the differential elastic scattering cross section, with $d\sigma_{\rm exp}/dq^2=\exp(A_{\rm exp}-B_{\rm exp}q^2)$. The Coulomb and Coulomb-hadronic interference contributions to the scattering were included in the fit. $A$ and $B$ are the corresponding parameters in  fits which included the curvature parameters $C$ and $D$, with  $d\sigma/dq^2=\exp(A-Bq^2+Cq^4-Dq^6)$. These were calculated using the comprehensive eikonal fit to the high energy $pp$ and $\bar{p}p$ data in \cite{bdhh-eikonal}.   \label{table:fitparameters}}				 
\def\arraystretch{1.15}            % Make the space between rows in the Table,
\begin{center}	
\begin{ruledtabular}			  % 1.5 x bigger than the default spacing.
\begin{tabular}[b]{|c|c|c|r||c|c|c|c|c|c|}
$W$, {\rm GeV} & $A_{\rm exp}$  & $B_{\rm exp}$ & $\chi^2_{\rm ref\ \ }$ & $A\ $ & $B\ $ & $C\ $ & $D\ $ & $\chi^2\ $ & ${\rm d.o.f.}\ $\\
 & & GeV$^{-2}$ & & & GeV$^{-2}$ & GeV$^{-4}$ & GeV$^{-6}$ & &  \\
\hline
  8000 &  $ 6.284\pm 0.001\ $ & $19.599\pm 0.018\ $ & $53.94\ $ & $6.300\pm 0.001\ $ & $20.280\pm 0.018\ $ &  $7.955\ $ & $25.58\ $ & 25.73 & $16\ $ \\
  7000 & $ 6.158\pm 0.002\ $ & $19.593\pm 0.039\ $ & $49.07\ $  & $ 6.168\pm 0.002\ $ & $20.197\pm 0.039\ $ & $ 8.229 $  & $25.09\ $ &  25.90 & $16\ $ \\
  1800 & $5.607\pm 0.023\ $ & $16.306\pm 0.375\ $ & 1$5.47\ $  & $ 5.632\pm 0.021\ $ & $17.296\pm 0.372\ $ & $10.132\ $  & $21.76\ $ & 14.69  & $23\ $  \\
   52.8 & $4.525\pm 0.001\ $ & $12.845\pm0.058\ $ & $79.71\ $  & $ 4.527\pm 0.001\ $ & $13.163\pm 0.058\ $ & $\ 6.817\ $  & $10.09\ $  & 70.82   & $32\ $  \\
\end{tabular}
\end{ruledtabular}

\end{center}
\end{table}
\def\arraystretch{1}
 %%%%%%%%%%%%%%%%%%%%%%%%%%%%
 %

 We find from Table I that the inclusion of the $C$ and $D$ terms, without any adjustment, improves the fits relative to the simple exponential fits in every case as indicated by the changes in the $\chi^2$, substantially so for the more precise TOTEM and ATLAS data. We take this as strong evidence for the presence of curvature in the differential cross sections.

 With the unlikelihood of a purely exponential behavior established, we can eliminate that possibility and apply the sieve procedure of Block \cite{sieve} to the favored model with curvature in $d\sigma/dq^2$. This procedure allows us to identify and eliminate possible outlying datum points relative to the behavior allowed in model.  The details of the sieve procedure are given in the appendix to \cite{sieve}.

 We  used this procedure with $\Delta_{\rm max}$ = 6 \cite{sieve} and found 1 outlier in the TOTEM data, 1 in the ATLAS data, none in the E710 data, and 2 in the ISR data. Eliminating those points led to substantial reductions in the $\chi^2$ for those fits, with only very small changes in the fitted parameters, all well within the uncertainties given in Table I. However, since we have not included systematic uncertainties in our analysis, only the statistical errors in the data, we will not use this refinement here and will use the parameters in Table I in the following.

\begin{figure}[htbp]
\includegraphics{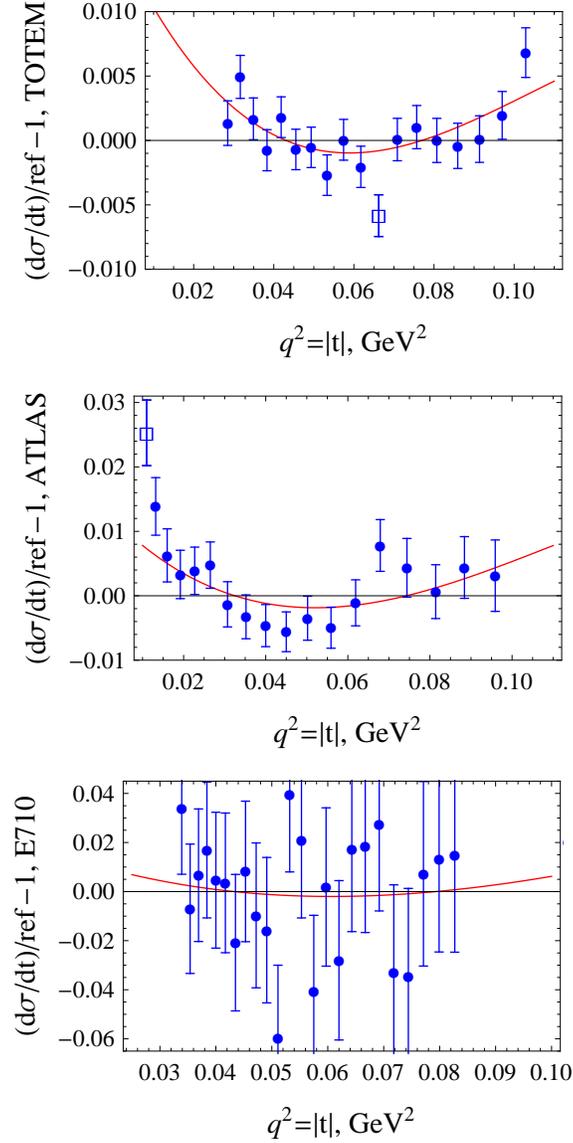}
\caption{Plots of the normalized differences $\left(d\sigma/dq^2-d\sigma_{\rm exp}/dq^2\right)\big/d\sigma_{\rm exp}/dq^2$ for our fits to the TOTEM \cite{totem2015}, ATLAS \cite{atlas2014}, and E710 \cite{E710_B2,E710lumin-ind,E710pbar-p-final} data at 8000, 7000, and 1800 GeV, respectively. The reference cross section was a purely exponential fit to the data with $d\sigma_{\rm exp}/dq^2=e^{A-B q^2}$, and corresponds in the figure to the horizontal line at 0. The final fits (solid red curves) included the curvature terms $+C q^4$ and $-D q^6$ in the exponent with the values of $C$ and $D$ at $q^2=0$ taken from the eikonal model in \cite{bdhh-eikonal};  $A$ and $B$ were determined in the fit. Only the quoted statistical uncertainties were used in the fit; these are shown in the figure. The points at 7000 and 8000 GeV identified as potential outliers in a sieve analysis are indicated by large open symbols; these were used in the fit.}
\label{fig:ratioTOTEM_ATLAS_E710}
\end{figure}

  We show our results for the higher energy data in a different way in \fig{fig:ratioTOTEM_ATLAS_E710}. There we plot the difference between cross sections $d\sigma/dq^2$ calculated for the final fit with the curvature effects included, and the simple exponential fit, all divided by the the cross section calculated for the exponential fit. The effects of the curvature terms are clearly evident in the theoretical curves, as is the improvement in the fits relative to the data when these are included. All the datum points in the $q^2$ intervals used  are shown. For reference,  the points identified as potential outliers at 7000 and 8000 GeV  through the sieve procedure are distinguished by large open symbols, but, as noted, these points were still used in making our fits and have only a minor effect on the results.

%%%%%%%%%%%%%%%%%%%%%
%%%%%% FIG 6  %%%%%

%%%%%%%%%%%%%%%%%%%%%
%

%%%%%%%%%%%%%%%%%%
%%%%%%%%%%%%%%%%%%

\subsection{Changes in the total cross sections and slope parameters \label{subsec:changes-in-B-sigma}}

%%%%%%%%%%%%%%%%%%

In most analyses of high-energy $pp$ and $\bar{p}p$ elastic scattering, the hadronic part of the near-forward differential cross section $d\sigma(W,q^2)/dq^2$ has been approximated as the exponential $\exp(A_{\rm exp}-B_{\rm exp}\,q^2)$.  The forward cross section $d\sigma(W,0)/dq^2$  in mb/GeV$^2$ is then just $\exp(A_{\rm exp})$, while the forward slope parameter is $B_{\rm exp}$. With the curvature effects included, $A_{\rm exp}\rightarrow A$ and $B_{\rm exp}\rightarrow B$ as in Table I.

The forward cross section is of particular interest since
\be
\label{sigma_of_0}
\frac{d\sigma}{dq^2}(W,0) = \frac{(1+\rho^2)}{16\pi}\sigma_{\rm tot}^2,
\ee
where $\rho$ is the ratio of the real to the imaginary part of the forward scattering amplitude, measurable through Coulomb-hadronic interference effects, and $\sigma_{\rm tot}$ is the total cross section.  This relation provides a measurement of $\sigma_{\rm tot}$ if $\rho$ and the forward cross section are known.

Using the results in Table I, we find the fractional changes in the slope parameter and total cross sections given in Table II. We find that the changes in the slope parameter relative to a purely exponential fit are quite significant at all the energies considered, with the final value of $B$ always several percent larger than the initial value obtained for an exponential fit in the range of energies shown. This is to be expected. The net curvature  corrections to the cross section are positive for the values of $q^2$ in question. This causes the actual differential cross section to curve upward away from the exponential fit as $q^2$ increases, and reduces the average slope found in the exponential fit.

The expression in \eq{Blocal} gives the estimate
\be
\label{B/Blocal}
B/B(q_0^2) \approx 1 +\left(2Cq_0^2-3Dq_0^4\right)/B(q_0^2)
\ee
for the fractional change in the slope in the extrapolation from $q_0^2$ to $q^2=0$. If we identify $B(q_0^2)$ with the slope $B_{\rm exp}$ found in the exponential fit, we find that $q_0^2$ should be about  4/10 of the way along the $q^2$ interval to match the ratios in Table II, that is, at $q_0^2\approx0.6q^2_{\rm min}+0.4q^2_{\rm max}$. This shift to  a point below the center of the interval is again to be expected because the cross section curve steepens as $q^2$  decreases, shifting the point at which $B_{\rm exp}$ and the local slope---the slope of the tangent curve----match to smaller $q^2$.

%%%%%%%%%%%%%%%%%%%%%%
%%%%%%% TABLE 2 %%%%%%%%%%

 \begin{table}[ht]                   % Use "table" environment, but also ruled tabular, tabular
 \caption{\protect  The fractional changes in the forward differential cross section $d\sigma/dq^2|_{q^2=0}=\exp{A}$ and the slope parameter $B$ obtained in fits to the data at 8000 \cite{totem2015}, 7000 \cite{atlas2014}, 1800 \cite{E710_B2}, and 52.8 GeV \cite{ISR1} over the $q^2$ intervals shown when the curvature terms $C$ and $D$ are included, relative to purely exponential fits. The values of $C$ and $D$ were calculated using the eikonal fit to the high energy $pp$ and $\bar{p}p$ data in \cite{bdhh-eikonal}. The final columns show the $\rho$ values used in converting the forward cross section to the total cross section $\sigma_{\rm}$, and the final results for the latter with their purely statistical uncertainties.
 \label{table:changes}}	
\def\arraystretch{1.15}          % increase spacing between lines for clarity
\begin{center}	
\begin{ruledtabular}			
\begin{tabular}[b]{|c|c|c|c|c|c|c|}   % | bar for vertical line, c center, d center on decimal (only one, no math mode), r, l right, left
$W$, GeV & $q^2$ range$\ \ $ &  $\exp{(A)}/\exp(A_{\rm exp})\  $ & $B/B_{\rm exp}\ $ &  $ d\sigma/dq^2|_{q^2=0}\ $ & $\rho\ \  \ $ & $\sigma_{\rm tot}\ \ $\\
 & & GeV$^2$   & GeV$^{-2}$ & mb/GeV$^2$ & & $ {\rm  mb}\ \ $ \\
\hline
8000 & 0.027--0.103 $\ $&  $ 1.016\pm 0.001\  $ & $ 1.035\pm 0.001\ $ & $ 540.8 \pm  0.10\ $ &  $ 0.131\  \ $ & $ 102.0 \pm 0.01\ $   \\
7000 & 0.011--0.096 $\ $& $ 1.010\pm  0.002\ $ & $ 1.031\pm  0.003\ $ & $  477.2\pm 0.15\ $ & $ 0.133\  \ $ &  $  95.8\pm 0.02 \  $ \\
1800 & 0.034--0.103 $\ $ & $ 1.025\pm 0.021\  $ & $ 1.061\pm  0.024\ $ & $ 279.2\pm 1.1\ \ $ & $ 0.144\ \  $ & $ 73.2\pm 0.14\  $ \\
52.8  & 0.001--0.056 $\ $ & $ 1.002\pm 0.000 \  $ & $  1.025\pm 0.006\ $ & $ 92.5\pm 0.03 $ & $ 0.073\ \ $ & $ 42.4 \pm 0.01\ $  \\
 \end{tabular}
\end{ruledtabular}

\end{center}
\end{table}
\def\arraystretch{1}
 %%%%%%%%%%%%%%%%%%%%%%%%%%%%
 %

\begin{figure}[htbp]
\includegraphics{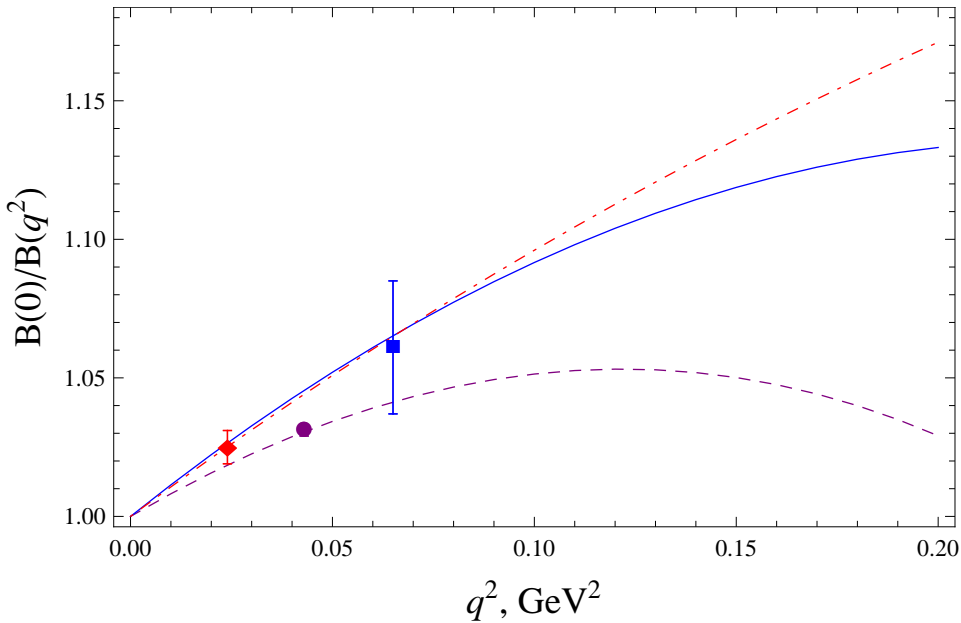}
\caption{Plots of the ratio $B(0)/B(q_0^2)$ for $W=52.8$ GeV (red dot-dashed curve), 1800 GeV (solid blue curve), and  7000 GeV (dashed purple curve).  The points shown for $W=52.8$ GeV (red diamond), 1800 GeV (blue square), and 7000 GeV (purple dot)  are the ratios obtained using the fitted values of $B(0)$ and the calculated local slopes $B(q_0^2)$ in the eikonal model of \cite{bdhh-eikonal} at the point $q_0^2=0.6 q_{\rm min}^2+0.4 q_{\rm max}^2$ in the $q^2$ range of the data.  }
\label{fig:Bratios}
\end{figure}

 We plot the ratios $B/B_{\rm exp}$ from Table II at the points $q_0^2$ specified above in \fig{fig:Bratios}, along with the the ratio curves for $B/B(q^2)$ calculated using the eikonal model of \cite{bdhh-eikonal}.  The agreement is good. We conclude that changes of several percent in the values of the forward slope parameter $B$ relative to the values determined in an exponential fit are to expected, with the magnitudes of the changes dependent on both the energy and the interval in $q^2$ used in the fit.

%%%%%%%%%%%%%%%%%%%%%
%%%%%% FIG 7  %%%%%

%%%%%%%%%%%%%%%%%%%%%
%

Our final value of $B$ at 8000  GeV in Table I agrees very well with the value obtained by the TOTEM  experiment in an analysis which included a fit to the $C$ and $D$ parameters, $B=20.14\pm 0.15$ GeV$^{-2}$ \cite{totem2015}. Our values of $B$ at 7000, 1800, and 52.8 GeV are higher than the experimental values obtained with purely exponential fits to the hadronic parts of the differential cross sections, but are reproduced within the experimental uncertainties by multiplying the latter by the factors $B/B_{\rm exp}$ in Table II.

We see from Table II that the changes in the forward differential cross sections, given by the ratios $\exp{(A)}/\exp{(A_{\rm exp})}$, are considerably smaller than the changes in the slope parameter $B$. This is again an effect of the curvature: the exponential least-squares fit must cut across the positively curved data in order to minimize the total $\chi^2$. This increases the value of $A_{\rm exp}$, reducing its difference from $A$. This effect is evident in figure \fig{fig:ratioTOTEM_ATLAS_E710}, where the exponential fit corresponds to the horizontal curve at zero.

The changes in the total cross sections, which appear squared in \eq{sigma_of_0}, are only half those in the forward cross sections. We give the resulting cross sections in Table II for the values of the ratio $\rho$ of the real to the imaginary part of the forward scattering amplitudes determined in  \cite{bdhh-eikonal}. The uncertainties listed for the cross sections are only statistical.

Our results for the total cross sections agree well with the published results: $101.9\pm 2.1$ mb for the TOTEM \cite{totem2015}, $95.35\pm 1.30$ mb for ATLAS \cite{atlas2014}, $72.1\pm 3.3$ mb for E710 \cite{E710lumin-ind}, and $43.38\pm 0.15$ mb for the ISR \cite{ISR1} experiments. With the expected small curvature corrections included, these become  $96.3\pm 1.30$ mb, $73.9\pm 3.3$ mb, and $43.47\pm 0.154 $ mb for ATLAS, E710, and the ISR, respectively. The TOTEM analysis already included curvature.

%%%%%%%%%%%%%%%%%%%%%%
%%%%%%%%%%%%%%%%%%%%%%

\section{Conclusions \label{sec:conclusions}}

We have investigated the effects of curvature on the $pp$ and $\bar{p}p$ cross sections at high energies ($>30$ GeV) using an eikonal model fitted to the combined data on $\sigma_{\rm tot}$, $\sigma_{\rm elas}$, $\sigma_{\rm inel}$, $\rho$, and the forward slope parameter $B$. We find that the effects are small but significant, leading to changes in $B$, the forward cross section $d\sigma(W,0)/dq^2$, and through the latter to $\sigma_{\rm tot}$. The changes to $B$ in particular are well outside the quoted experimental uncertainties.

It is our conclusion that the effects of curvature in the small-$t$ differential cross sections should be included in fits to  new data, and in reanalyses of existing data. While the existing data are generally not precise enough to determine the curvature terms in $\ln(d\sigma/dt)$ directly, we find that a hybrid approach in which those small terms are taken from the eikonal model and only the forward cross section and the slope parameter are adjusted at small $t$ leads to improvements relative to the existing results. We note also that the fitting should generally be restricted to the range $|t|\lesssim 0.1$ GeV$^2$ as curvature effects become large and increasingly uncertain at larger values of $|t|$.

%%%%%%%%%%%%%%%%%%%%%%
%%%%%%%%%%%%%%%%%%%%%%

\begin{acknowledgments}

The authors wish to thank Profs.\  Leo Stodolsky and Thomas J.\ Weiler for their stimulating input in early discussions of this work.
M.M.B., L.D., and F.H.\  would  like to thank the Aspen Center for Physics for its hospitality and for its partial support of this work under NSF Grant No. 1066293. F.H.'s research was supported in part by the U.S. National Science
Foundation under Grants No.~OPP-0236449 and PHY-0969061 and by the
University of Wisconsin Research Committee with funds granted by the Wisconsin Alumni Research
Foundation.   P.H.\ would like to thank the Towson University Fisher College of Science and Mathematics for support.

\end{acknowledgments}

%%%%%%%%%%%%%%%%%%%%%%
%%%%%%%%%%%%%%%%%%%%%%

\appendix

\bibliography{small_x_references-4-16}

\end{document}